 \documentclass[twocolumn,showpacs,preprintnumbers,amsmath,amssymb]{revtex4}

\usepackage{graphicx}
\usepackage{dcolumn}
\usepackage{bm}
\usepackage{color}

\begin{document}

\title{Chiral geometry of higher excited bands
in triaxial nuclei with particle-hole configuration}

\author{Q. B. Chen}
 \affiliation{ School of Physical Science and Technology, Southwest
 University, Chongqing 400715, China}
\author{J. M. Yao}\email{jmyao@swu.edu.cn}
 \affiliation{ School of Physical Science
and Technology, Southwest University, Chongqing 400715, China}
\author{S. Q. Zhang}
 \affiliation{State Key Laboratory of Nuclear Physics and Technology,
  School of Physics, Peking University, Beijing 100871, China}

\author{B. Qi}
\affiliation{Shandong Provincial Key Laboratory of Optical Astronomy
and Solar-Terrestrial Environment, School of Space Science and
Physics, Shandong University at Weihai, Weihai 264209, China}

 \date{\today}

\begin{abstract}
The lowest six rotational bands have been studied in the
particle-rotor model with the particle-hole configuration $\pi
h^1_{11/2}\otimes\nu h^{-1}_{11/2}$ and different triaxiality
parameter $\gamma$. Both constant and spin-dependent variable
moments of inertial (CMI and VMI) are introduced.  The energy
spectra, electromagnetic transition probabilities, angular momentum
components and $K$-distribution have been examined. It is shown
that, besides the band 1 and band 2, the predicted band 3 and band 4
in the calculations of both CMI and VMI for atomic nuclei with
$\gamma=30^\circ$ could be interpreted as chiral doublet bands.
\end{abstract}

\pacs{21.60.Ev, 21.10.Re, 23.20.Lv}  \maketitle


Chirality is a subject of general interest in molecular physics,
elementary particle physics, and optical physics. In atomic nuclear
physics, the occurrence of chirality was originally suggested in
1997 by Frauendorf and Meng in the particle-rotor model (PRM) and
tilted axis cranking (TAC) approach for triaxially deformed
nuclei~\cite{M1997}. The predicted patterns of spectra exhibiting
chirality were experimentally observed in 2001~\cite{K2001}. Since
then, the investigation of chiral symmetry in atomic nuclei has
become one of the most hot topics in nuclear physics. Hitherto, more
than 20 candidate chiral doublet bands in odd-odd nuclei are
proposed in the $A\sim100$, $A\sim130$, and $A\sim190$ mass regions.
In addition, a few more candidates with more than one valence
particle and hole were also reported in odd-$A$ and even-even
nuclei. For a review, see e.g.~\cite{M2008}. Even though there are
many candidate chiral nuclei, the interpretation of the observed
pair of near degenerate $\Delta I=1$ bands with the same parity as
the chiral doublet bands is still an open question, accompanied with
several competitive mechanisms~\cite{Meng09}.

On the theoretical side,  chiral doublet bands were first predicted
by particle-rotor model (PRM) and tilted axis cranking (TAC) model
for triaxially deformed nuclei~\cite{M1997}. Later on, numerous
efforts have been devoted to the development of TAC
methods~\cite{D2000,Olbratowski04,Olbratowskiprc} and PRM
models~\cite{Peng03,Koike04,Wang07,Z2007} to describe chiral
rotation in atomic nuclei. The characters of chiral doublet bands
have been
examined~\cite{Koike03,Petrache2006,Mukhopadhyay07,Wang2007,Qi2009}.
Recently, the PRM has been extended to the case of many particles
and/or holes couples to a triaxially deformed core~\cite{Q2009} that
allows for the study of more general chiral rotating nuclei. As the
counterpart, the interacting boson-fermion-fermion model has also
been used to study the chiral doublet bands~\cite{Tonev06,Tonev07}.

In Ref.~\cite{M2006}, the possible existence of multi-chiral doublet
bands (M$\chi$D) in a single-nucleus $^{106}$Rh has been proposed
based on the triaxial relativistic mean-field calculations and been
confirmed later in the similar calculations but with time-odd
components~\cite{Y2009}. Several minima with large triaxial
deformation but with different high-$j$ proton-particle and
neutron-hole configurations, which are favorable for nuclear
chirality, were found in the calculated potential energy surfaces of
the rhodium isotopes $^{104, 106, 108, 110}$Rh~\cite{P2008}.

In comparison with the M$\chi$D that differ from each other in the
triaxial deformations and multiparticle configurations, more than
one pair of chiral doublet bands may also exist in a single nucleus
with the same particle-hole configuration, i.e., not only the yrast
and yrare bands but also two higher excited bands might be chiral
partners. In the PRM with either a rigid or a soft triaxial core,
the properties of chiral bands, including the yrast and yrare bands
as well as the higher excited bands in an odd-odd nucleus were
calculated~\cite{Droste09}. It has been shown that the properties of
the two higher excited bands (bands 3 and 4), including the reduced
probabilities of E2 and M1 transitions between two states with
$\Delta I=1$ and staggering patterns, are similar to those of the
yrast and yrare bands. It gives us a hint that bands 3 and 4 could
be chiral partners as well~\cite{Droste09}. These facts motivate us
to make more detailed theoretical studies for the higher excited
bands of chiral rotating nuclei. Particular attention will be paid
to examining their chiral geometry.

The PRM, in which the total angular momentum is a good quantum
number, has made great success in the investigation of chiral
rotating nuclei in different mass region. Therefore, in this brief
report, we would like to adopt this model to make a further study
for the lowest six rotational bands (bands 1-6), searching for the
higher chiral doublet bands.

The detailed formalism for PRM can be found in
Refs.~\cite{M1997,Peng03,Koike04}. In the calculations, we take the
symmetric particle-hole configuration $\pi h_{11/2}\otimes \nu
h_{11/2}^{-1}$ corresponding to the $A\sim130$ mass region and the
triaxial deformation parameter $\gamma=30^\circ$. In the calculation
of electromagnetic transitions, the intrinsic quadrupole moment
$Q_0=(3/\sqrt{5\pi})R_0^2Z\beta$ is taken to be 3.5 eb. The
gyromagnetic ratios $g_R=Z/A=0.44$, and $g_p=1.21$, $g_n=-0.21$ are
adopted as Ref.~\cite{Wang07}. Both constant $\mathcal{J}_0$ (CMI)
and spin-dependent variable $\mathcal{J}(I)$ (VMI) moment of inertia
are introduced to simulate the rigid and variable cores, where the
VMI~\cite{Mariscotti69} is taken the same form as
Refs.~\cite{Zeng85,Chen2006}
 \begin{equation}\label{MOI}
 \mathcal{J}=\mathcal{J}_0\sqrt{1+bI(I+1)},
 \end{equation}
 with $\mathcal{J}_0=30$ MeV$^{-1} \hbar^2$ and $b=0.01$.


\begin{figure}[]
\hspace{-1.0 cm}
\includegraphics[width=8cm]{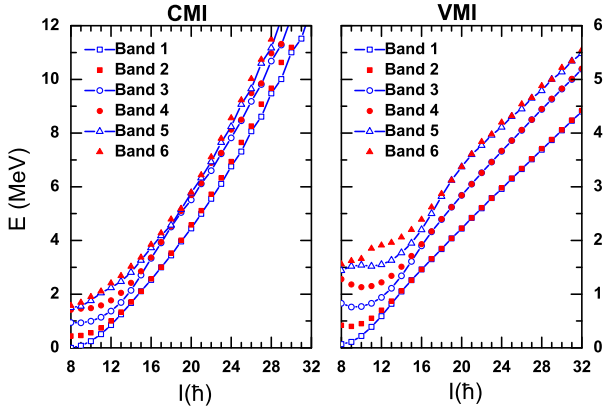}
\caption{The six lowest energy levels as a function of spin,
calculated by the PRM with the CMI (left panel) and VMI (right
panel). The corresponding grouped bands are labeled as band 1-6 in
accordance with the increasing energy of the state $8^+_\alpha$.}
\label{fig1}
\end{figure}

Figure~\ref{fig1} displays the six lowest energy levels
$I_{\alpha=1,6}$ as a function of spin $I$, calculated by the PRM
with both CMI and VMI. All the calculated levels have been grouped
into six bands which are labeled as bands 1-6 in accordance with the
order of the energies of  the states $8^+_\alpha$.

As shown in Fig.~\ref{fig1}, the degeneracy is found not only
between band 1 and band 2, but also between band 3 and band 4 in
some spin region. Quantitatively, the energy difference between the
band 3 and band 4 in $I\geq 16\hbar$ region is within 310 keV in the
calculation with CMI, and 30 keV in the calculation with VMI.
Similar degeneracy with relative larger energy difference is
observed between band 5 and band 6. It is noted that the excitation
energy $E(I)$ from the calculation of VMI becomes approximately
linear spin-dependent as the spin $I\gg 10\hbar$, in which case, the
collective rotation of core becomes dominate with moment of inertia
$\mathcal{J}\simeq3I$~MeV$^{-1} \hbar^2$.

\begin{figure}[]
\begin{center}
\includegraphics[width=7cm]{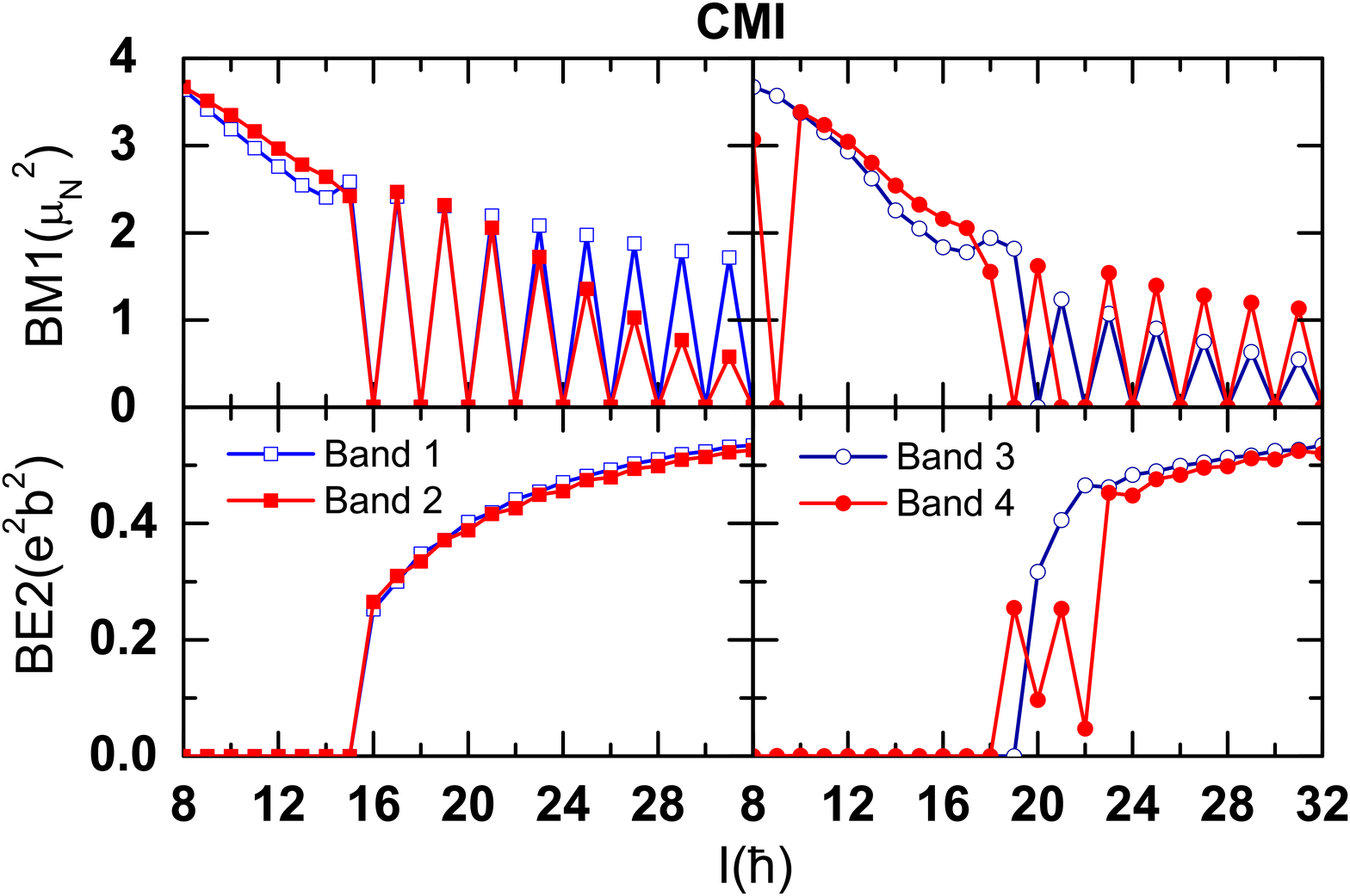}
\includegraphics[width=7cm]{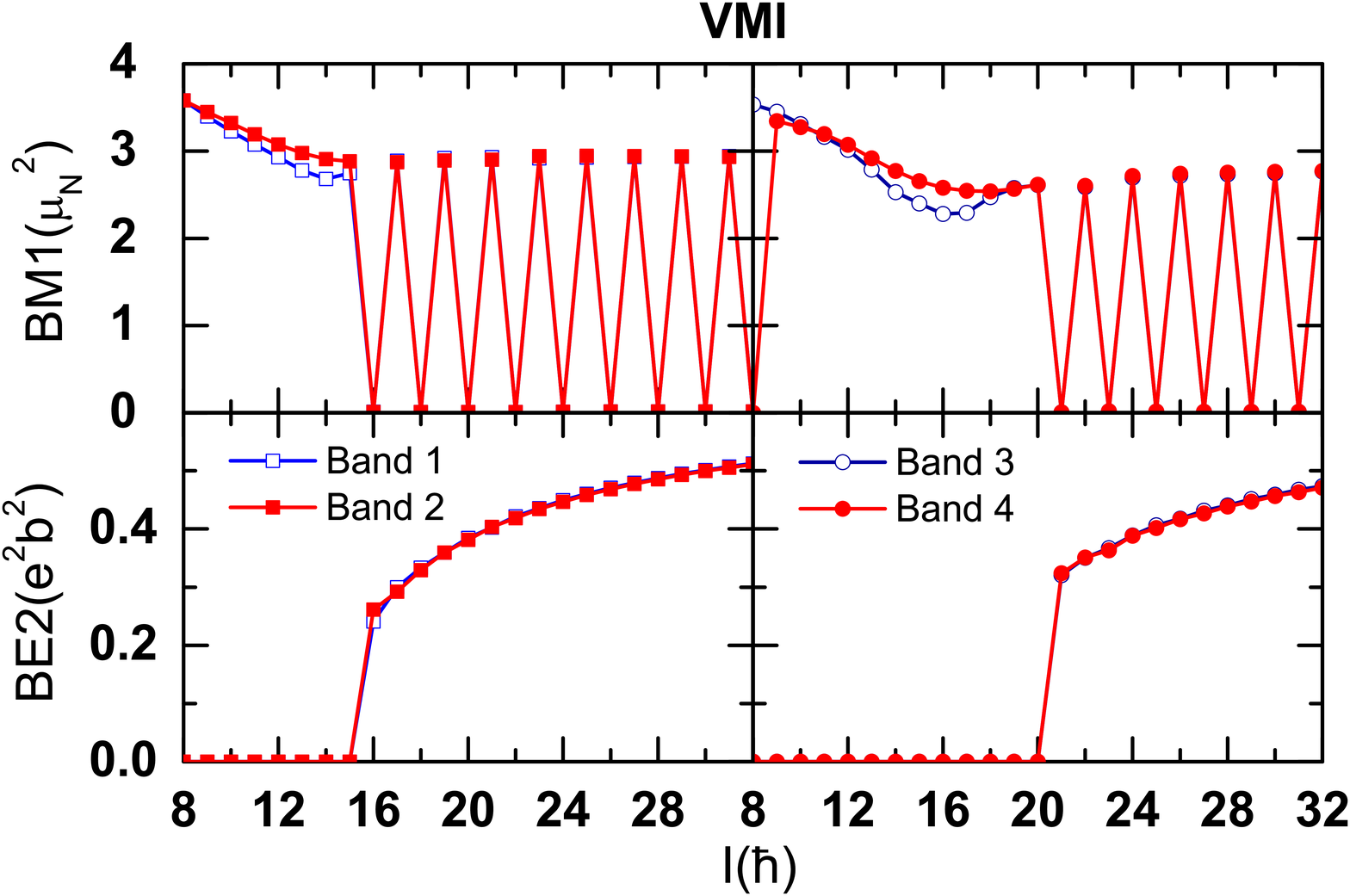}
\end{center}
\caption{The intraband electromagnetic transition probabilities
$B(M1)$ and $B(E2)$ of band 1-4 as functions of the total spin,
calculated in PRM with the CMI (upper panel) and VMI (lower panel).}
\label{fig2}
\end{figure}

Besides the small energy difference, the selection
rule~\cite{Koike04} and the similarity of the electromagnetic
transition probabilities are remarkable characters for chiral
doublet bands. The intraband electromagnetic transition
probabilities $B(M1;I\to I-1)$ and $B(E2;I\to I-2)$ of bands 1-4 in
the PRM calculations with the CMI and VMI are plotted in
Fig.~\ref{fig2}. As discussed in Refs.~\cite{Koike04,Qi2009}, the
band 1 and band 2 at $I=15 \hbar$, as well as the band 3 and band 4
at $I=17, 18, 19 \hbar$, from the calculations with CMI have been
exchanged to ensure that the bands are organized based on $B(E2)$.
Similarly, for the calculations with VMI, one should exchange the
band 1 and band 2 at $I=17, 19, 21\hbar$, as well as the band 3 and
band 4 at $I=19, 21\hbar$. It is shown that the intraband
electromagnetic transition probabilities $B(M1)$ and $B(E2)$ of the
band 3 and band 4 are quite similar in the calculations with CMI and
almost exactly the same in the calculations with VMI. Moreover,
similar behaviors of $B(M1)$ and $B(E2)$ as functions of spin in
band 3 and band 4 with those in band 1 and band 2 are observed. The
obvious difference is the spin value where the odd-even spin
staggering in $B(M1)$ or $B(M1)/B(E2)$ occurs. The above phenomena
indicate that the static chirality may form between band 3 and band
4 in spin $I\geq20\hbar$ region.

It is noted that an evident difference between the calculations with
the CMI and VMI is the amplitudes of odd-even spin staggering in
$B(M1)$ values, which are gradually decreasing with the increasing
of spin in the calculations of CMI. While, in the calculations of
VMI, the amplitudes of odd-even spin staggering are almost
spin-independent. It implies that the calculations of VMI would
present a little different chiral geometry from that given by the
calculations of CMI.


\begin{figure}[]
\includegraphics[width=6cm]{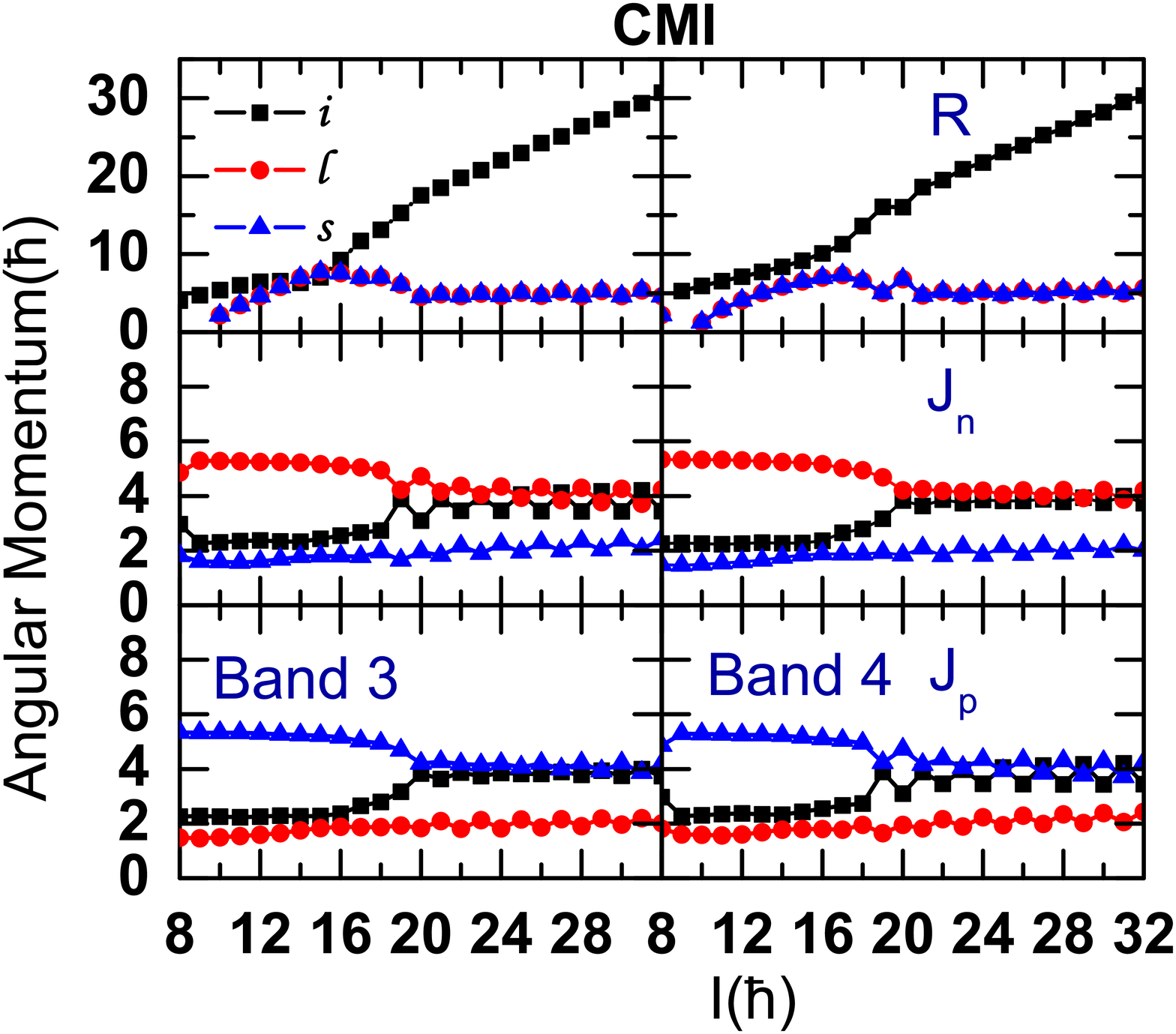}
\includegraphics[width=6cm]{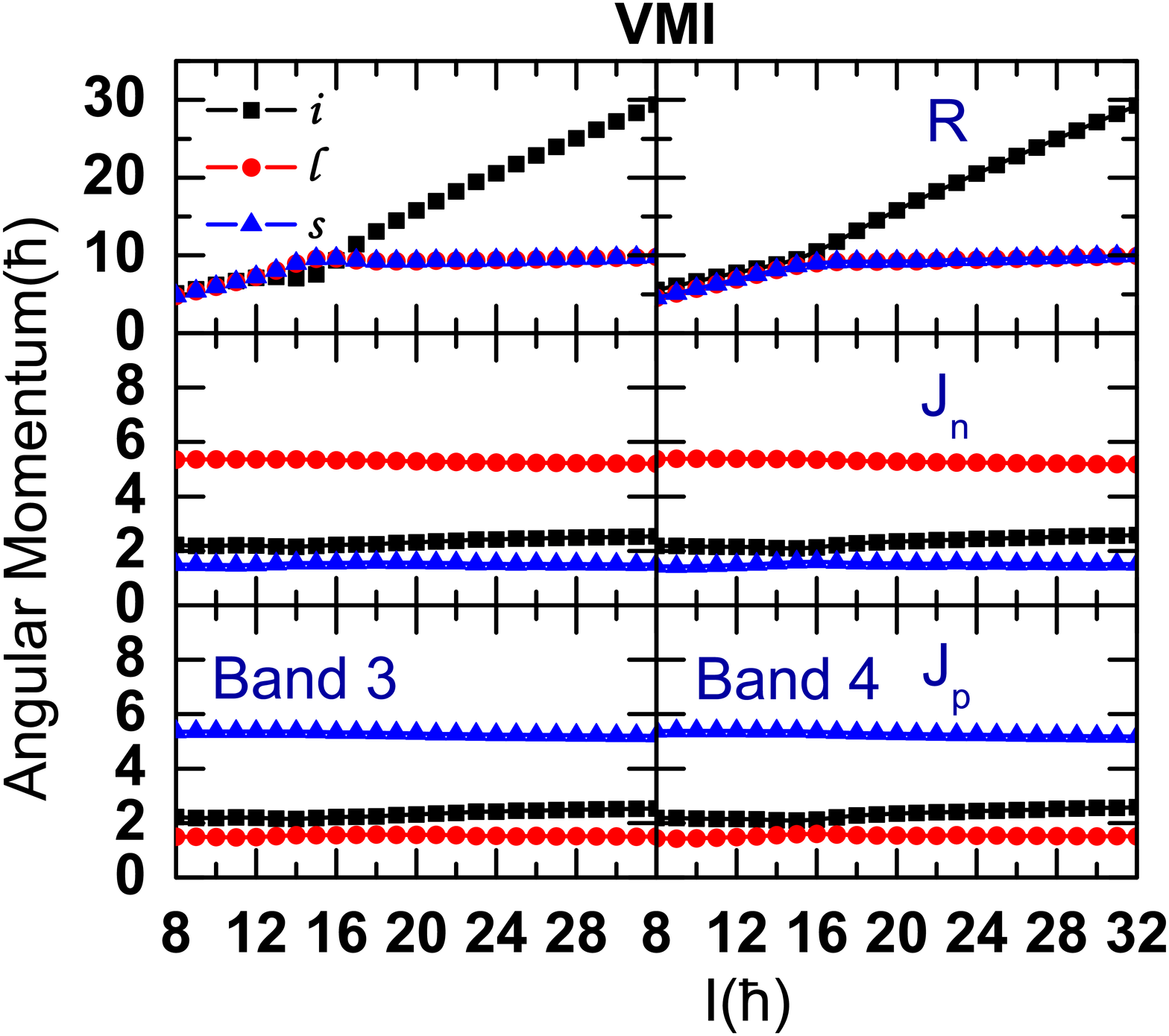}
\caption{Components along the intermediate (i, squares), short (s,
triangles), and long (l, circles) axes of the angular momenta of the
core $R_k=\langle \hat{R}_k^2 \rangle^{1/2}$, valence neutron
$J_{nk}=\langle \hat{j}_{nk}^2\rangle^{1/2}$, and valence proton
$J_{pk}=\langle \hat{j}_{pk}^2 \rangle^{1/2}$ for the band 3 and
band 4 in PRM with both CMI (upper panel) and VMI (lower panel). }
\label{fig3}
\end{figure}
%
 \begin{figure}[]
\begin{center}
\includegraphics[width=6cm]{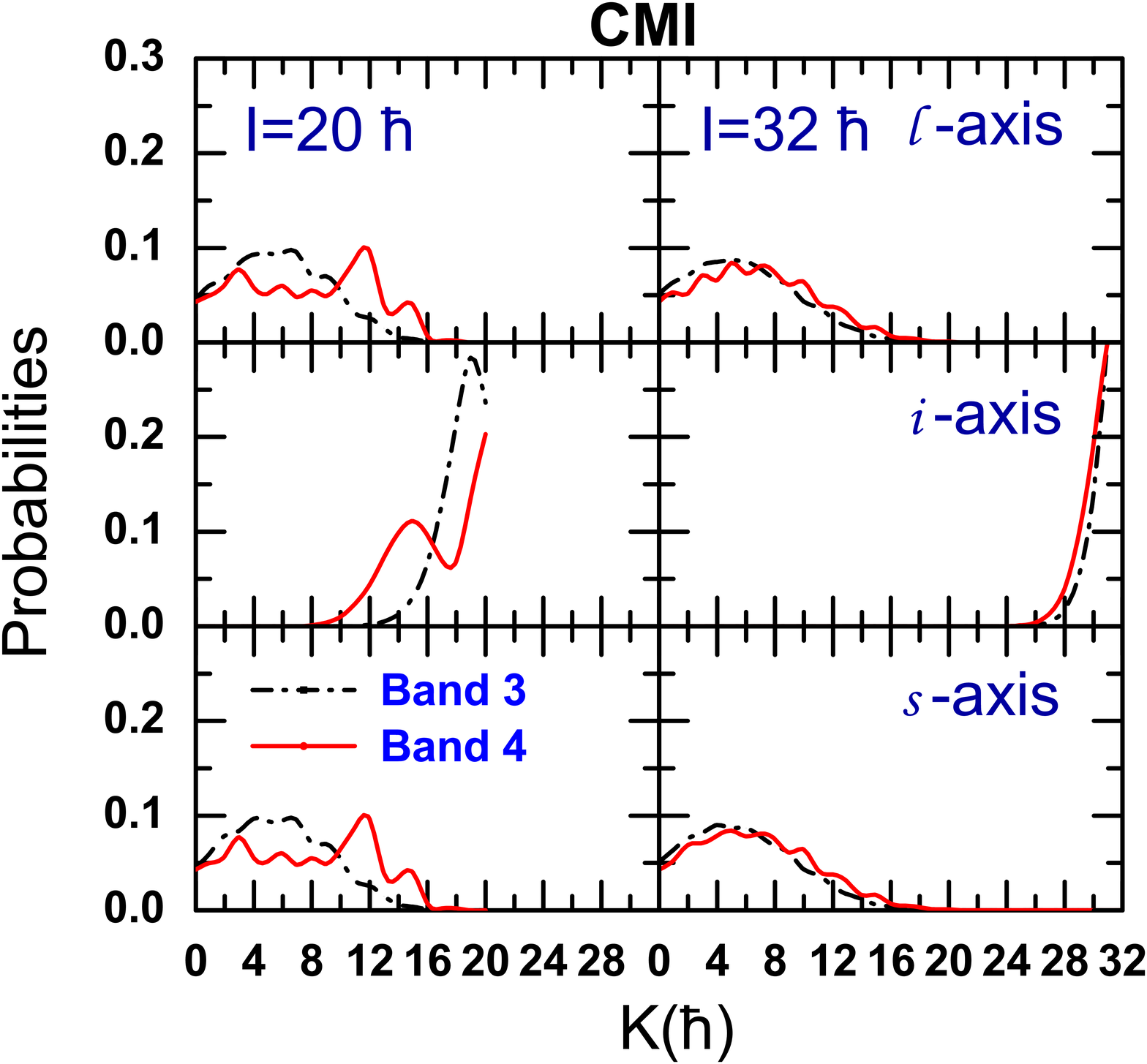}
\includegraphics[width=6cm]{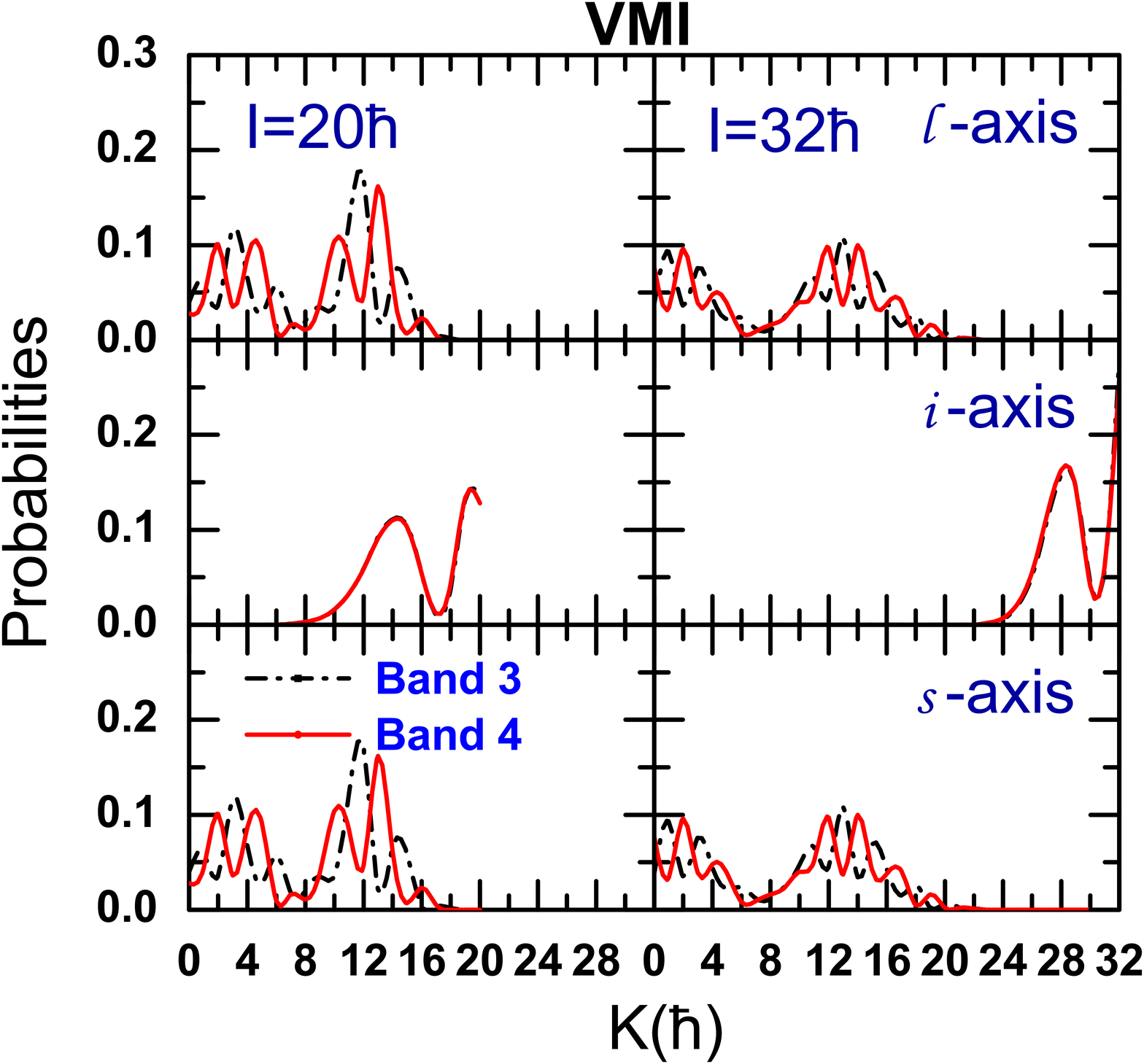}
 \caption{Probability distributions for projections $K$ of total
 angular momentum on the long (l), intermediate (i), and short (s)
axes for the band 3 and band 4 in PRM with both CMI (upper panel)
and VMI (lower panel).}
 \label{fig4}
 \end{center}
\end{figure}

In addition, we have calculated the electromagnetic transition
probabilities in band 5 and band 6. The intraband transitions
$B(M1)$ and $B(E2)$ of band 5 and band 6 are quite different,
irrespective of what $\gamma$ value being used.

The ideal chiral picture in atomic nuclei is formed by three
mutually perpendicular angular momenta, i.e., the collective angular
momentum of triaxially deformed core with $\gamma=30^\circ$ favors
alignment along the intermediate axis, whereas the angular momentum
vectors of the high-$j$ valence particles (holes) favor alignment
along the nuclear short (long) axis. Therefore, to examine the
chiral geometry of band 3 and band 4, the rms values of the angular
momentum components for the core $R_k=\langle \hat{R}_k^2
\rangle^{1/2}$, the valence neutron $J_{nk}=\langle
\hat{j}_{nk}^2\rangle^{1/2}$, and the valence proton $J_{pk}=\langle
\hat{j}_{pk}^2 \rangle^{1/2}$ for the band 3 and band 4 from the PRM
calculations with both CMI and VMI are plotted in Fig.~\ref{fig3},
where the indices $k=i, l, s$ represent the intermediate(i),
short(s), and long(l) axes respectively.

As shown in the Fig.~\ref{fig3}, the calculations with CMI and VMI
both present the aplanar rotation for band 3 and band 4.
Specifically, the core angular momentum mainly aligns along the
intermediate axis in the calculations with both CMI and VMI due to
its largest moment of inertia. For the case of CMI, the $h_{11/2}$
valence neutron hole mainly aligns along the long axis at the
beginning of the bands. As the total angular momentum
$I\geq20\hbar$, it lies in the plane defined by the $l$ axis and $i$
axis. In the mean time, the $h_{11/2}$ valence proton lies in
another plane defined by the $s$ axis and $i$ axis. The three
angular momenta together form an aplanar geometry. Such a picture is
slightly different from that of VMI, in which case, the $h_{11/2}$
valence neutron hole mainly aligns along the $l$ axis, and the
valence proton mainly along the $s$ axis at all spin values.

Furthermore, it has been shown in Ref.~\cite{Q2009} that the static
chiral rotational bands (band 1 and band 2) have similar
$K$-distribution, $P^I_K=\sum_{\alpha}|C^I_{K\alpha}|^2$, where the
$C^I_{K\alpha}$ is the expansion coefficient of total wave function
on the strong coupling basis~\cite{Z2007}. Similar $K$-distribution
has been found between static chiral doublet bands. In
Fig.~\ref{fig4}, we display the $K$-distribution of $I=20\hbar$ and
$32\hbar$ states in band 3 and band 4 on the $l, i, s$ axes from the
PRM calculations with the CMI and VMI. In the calculations with the
CMI, the $K$-distribution of $I=20\hbar$ states is somewhat
different in band 3 and band 4. As the increasing of spin, the
$K$-distribution becomes similar, especially at $I=32\hbar$. In
comparison with the results of CMI, the $K$-distribution of band 3
and band 4 is similar even for the state with spin $I=20\hbar$. It
indicates that band 3 and band 4 have the characters of static
chirality in spin $I\geq20\hbar$ region. Moreover, the calculations
with the VMI present better chiral characters in band 3 and band 4
than those in the calculations with the CMI.

In addition, the properties of band 3 and band 4 from the PRM
calculations with different triaxiality parameter between
$15^\circ\leq \gamma \leq 30^\circ$ are also examined. We find that
the deviation of $\gamma$ from $30^\circ$ makes the static chirality
violated quickly in band 3 and band 4. It indicates that the chiral
geometry of band 3 and band 4 is much more sensitive to the triaxial
deformation than that of band 1 and band 2.


In summary, the lowest six bands have been studied in the
particle-rotor model with the configuration $\pi
h^1_{11/2}\otimes\nu h^{-1}_{11/2}$ corresponding to $A\sim130$ mass
region and different values of triaxiality parameter. Both constant
and spin-dependent variable moments of inertial have been
introduced. The energy spectra, electromagnetic transition
probabilities, angular momentum components and $K$-distribution have
been examined. All calculated quantities indicate that besides the
band 1 and band 2, the band 3 and band 4 could also form chiral
doublet bands in certain spin region ($I\geq20\hbar$) for atomic
nuclei with triaxiality parameter $\gamma=30^\circ$. However, the
chiral geometry of band 3 and band 4 has been found to be strong
sensitive to the $\gamma$ value. The calculations with the
spin-dependent variable moment of inertial have been shown to
present better chiral rotation characters of band 3 and band 4 than
that with the constant one.

 \begin{acknowledgments}
We thank Professor Jie Meng for useful discussion. This work was
partly supported by the Major State 973 Program 2007CB815000 and the
National Natural Science Foundation of China under Grant Nos.
10947013, 10975008 and 10775004, the Southwest University Initial
Research Foundation Grant to Doctor (No. SWU109011).
\end{acknowledgments}


\end{document}